\documentclass[showpacs]{revtex4} 
\usepackage{epsfig,amsmath,amssymb,graphicx,upgreek,textcomp}
\usepackage{natbib}
\usepackage[english]{babel} %

\begin{document}

\vspace{0mm}
\title{To the theory of phase transition of a binary solution into an
inhomogeneous phase} %
\author{Yu.M. Poluektov}
\email{yuripoluektov@kipt.kharkov.ua (y.poluekt52@gmail.com)} %
\affiliation{National Science Center ``Kharkov Institute of Physics and Technology'', 61108 Kharkov, Ukraine} %
\author{A.A. Soroka} %
\affiliation{National Science Center ``Kharkov Institute of Physics and Technology'', 61108 Kharkov, Ukraine} %

\begin{abstract}
In the framework of the theoretical model of the phase transition of
binary solutions into spatially inhomogeneous states proposed
earlier by the authors [1], which takes into account nonlinear
effects, the role of the cubic in concentration term in the
expansion of free energy was studied. It is shown that taking into
account the cubic term contributes to the stabilization of a
homogeneous state. The existence of two inhomogeneous phases in an
isotropic medium, considered in [1], proves to be possible only at
half the concentration of the solution. The contribution of
inhomogeneity effects to thermodynamic quantities is calculated.
Phase transitions from a homogeneous state and between inhomogeneous
phases are second-order phase transitions.
\newline%
\end{abstract}
\maketitle

\section{Introduction}\vspace{-0mm} 
In [1], the authors proposed a model for the phase transition of a
binary solution into an inhomogeneous state and studied the
thermodynamics of this process. This work was a development of the
approach proposed earlier in articles [2--5]. A feature of the model
considered in [1] is that it assumes, as is the case in real
conditions, a fixed average concentration. The free energy of the
solution is chosen as a polynomial of the fourth degree in the
concentration fluctuations. In [1] the authors analyzed in detail
the particular case when there is absent a cubic term in the
expansion of free energy, which can be realized in crystals of a
certain crystal symmetry [6]. It was shown that in this case the
solution, as a result of the loss of stability of a spatially
homogeneous state, can at an arbitrary concentration go either into
a state with a concentration wave ({\it W}\,-phase), or into a state
with stratification which is described by a solution in the form of
a "kink" ({\it K}\,-phase).

The purpose of this work is to study the influence of the term in
the free energy, which is cubic with respect to concentration
fluctuations, on the character of transition to an inhomogeneous
state. Within the framework of the model of an isotropic medium, it
is shown that taking into account the cubic contribution leads to
the stabilization of a homogeneous state. The transition to an
inhomogeneous state proves to be possible only at the half
concentration, when the cubic in concentration term vanishes. At
other concentrations there also exist coordinate-dependent solutions
in theory, but they do not satisfy the requirement of conservation
of the number of particles, and therefore are not realized
physically. The contribution of inhomogeneity to thermodynamic
characteristics of a binary solution is studied and it is shown that
transitions between phases are second-order transitions with a jump
in the heat capacity.

\section{Formulation of the task }\vspace{-0mm} %
We consider a solution of particles of two types, the number of
which is $N_1$ and $N_2$, occupying a constant volume $V$. In the
isotropic approximation and for a spatially homogeneous state, the
thermodynamic properties of such a solution can be described using a
free energy, which is a function of temperature $T$, volume $V$, and
depends on the number of particles of each component:
$F=F(T,V,N_1,N_2)$. Sometimes it is more convenient to proceed to
the description using the total number of particles $N=N_1+N_2$ and
the concentration $c_1\equiv\overline{c}=N_1/N$  of one of the
components. As is known [7], in a spatially homogeneous state the
free energy for the one-component case can be represented as
$F=N\varphi(n,T)$. In the case of two components, a similar
relationship will take the form $F=N\varphi(n,T,c)$, where $n=N/V$
is the total density of the number of particles, which will be
assumed to be constant. In a spatially inhomogeneous state, the
concentration $c=c({\bf r})$ will be considered as a continuous
function of spatial coordinates. In this case, the total free energy
will be written in the form
\begin{equation} \label{01}
\begin{array}{l}
\displaystyle{%
   \overline{F}=n\!\int \!\varphi\big(T,n,\overline{c}\,; \delta c({\bf r}), \nabla\delta c({\bf r})\big)\, d{\bf r}.    %
}
\end{array}
\end{equation}
Thermodynamic potential (1) is a potential with incomplete
thermodynamic equilibrium. The temperature and density in it are
assumed to be equilibrium and independent of the coordinates, and
the order parameter, in our case it is the deviation of the
concentration from its equilibrium average value %
$\delta c({\bf r})=c({\bf r})-\overline{c}$, should be found as a
result of varying the free energy (1) with respect to the
concentration from the condition $\delta \overline{F}=0$. This leads
to the equation
\begin{equation} \label{02}
\begin{array}{l}
\displaystyle{%
   \frac{\partial \varphi}{\partial\delta c}-\nabla \frac{\partial \varphi}{\partial\nabla\delta c}=0. %
}
\end{array}
\end{equation}
Usually, the number of particles in a system is assumed to be fixed,
since it is determined by the composition of a sample; therefore, we
will consider the average concentration $\overline{c}=N_1/N$ as a
fixed parameter. By virtue of the requirement of conservation of the
total number of particles, the following conditions must be
satisfied
\begin{equation} \label{03}
\begin{array}{l}
\displaystyle{%
  \frac{1}{V}\!\int \! \delta c({\bf r}; T,n,\overline{c})\, d{\bf r} =0,  \qquad   %
  \overline{c}=\frac{1}{V}\!\int \! c({\bf r}; T,n,\overline{c})\, d{\bf r}.   %
}%
\end{array}
\end{equation}
Solutions of equation (2), which depend on temperature, density and
the average concentration $\delta c=\delta c({\bf r};
T,n,\overline{c})$, should be substituted into the expression for
the free energy (1). As a result, we obtain the formula for the
equilibrium free energy
\begin{equation} \label{04}
\begin{array}{l}
\displaystyle{%
   F=n\!\int \!\varphi\big(T,n,\overline{c}\,; \delta c({\bf r}; T,n,\overline{c}), \nabla\delta c({\bf r}; T,n,\overline{c})\big)\, d{\bf r}.    %
}
\end{array}
\end{equation}
The total entropy of the solution is determined through the
derivative of the equilibrium free energy (4) with respect to
temperature at constant parameters $n$ and $\overline{c}$: %
\begin{equation} \label{05}
\begin{array}{l}
\displaystyle{%
   S=-\frac{\partial F}{\partial T}=-n\int\frac{\partial '\varphi}{\partial T}\,d{\bf r}.    %
}
\end{array}
\end{equation}
In this case, due to condition (2), only the explicit dependence of
$\varphi$ on temperature should be differentiated, which is marked
in (5) with a prime. The equilibrium total energy of the solution is
determined by the usual formula:
\begin{equation} \label{06}
\begin{array}{l}
\displaystyle{%
   E=F+TS=n\int\!\left(\varphi-T\frac{\partial '\varphi}{\partial T}\right)\!d{\bf r}.    %
}
\end{array}
\end{equation}

In the proposed model, we choose the free energy density as an
expansion in powers of the concentration fluctuation up to the
fourth power inclusively:
\begin{equation} \label{07}
\begin{array}{l}
\displaystyle{%
   \varphi=\varphi_0(T,n,\overline{c})+\frac{K}{2}\big(\nabla\delta c\big)^2 + a_2\frac{(\delta c)^2}{2}+a_3\frac{(\delta c)^3}{3}+a_4\frac{(\delta c)^4}{4}.  %
}
\end{array}
\end{equation}
The first term $\varphi_0(T,n,\overline{c})$ in (7) describes the
contribution of the homogeneous state to the total free energy. In
what follows, we will be interested in the effects associated with
the inhomogeneity of a state. The linear term in (7) drops out due
to the conservation condition (3). The coefficients in the expansion
(7) depend, generally speaking, on the temperature, density, and
also the average concentration $a_i=a_i(T,n,\overline{c})$, $K=K(T,n,\overline{c})$. %
To ensure the stability of the homogeneous state at high
temperatures, the coefficients $K$ and $a_4$ should be considered
positive, and the coefficient $a_3$ can be either positive or
negative. Suppose that the coefficient $a_2$ changes sign at a
certain temperature $T_p$, becoming negative at $T<T_p$. This
temperature is also a function of the average concentration and
density $T_p=T_p(\overline{c},n)$. As will be seen later, one can
use the linear approximation $a_2=a_0\tau,\,\, \tau\equiv(T-T_p)\big/T_p\,\,(a_0>0)$. %
For the free energy density of the form (7), formula (2) leads to
the equation
\begin{equation} \label{08}
\begin{array}{l}
\displaystyle{%
   K\Delta\delta c=a_2\delta c+a_3(\delta c)^2+a_4(\delta c)^3.  %
}
\end{array}
\end{equation}
Among all possible solutions of this nonlinear equation, only those
ones are physically realizable for which $0\leq c({\bf r})\leq 1$.

The explicit form of the phenomenological parameters entering into
expansion (7) and their dependence on the average concentration,
temperature, and density can be found using the well-known
representation of the free energy [6,\,8], which in the considered
case of an isotropic medium can be represented as
\begin{equation} \label{09}
\begin{array}{l}
\displaystyle{%
   \overline{F}=\frac{n^2}{2}\!\int \! d{\bf r}d{\bf r}' V\big(|{\bf r}-{\bf r}'|\big)c({\bf r})c({\bf r}') + %
   nT\!\int \! d{\bf r}\Big[c({\bf r})\ln c({\bf r}) + \big(1-c({\bf r})\big)\ln \big(1-c({\bf r})\big) \Big],  %
}
\end{array}
\end{equation}
where $V\big(|{\bf r}-{\bf r}'|\big)$ is the interaction potential
between particles. Expanding (9) in terms of small fluctuations %
$\delta c({\bf r})=c({\bf r})-\overline{c}$\, and comparing with
(4),\,(7), we obtain the following expressions for the expansion
coefficients:
\begin{equation} \label{10}
\begin{array}{ccc}
\displaystyle{%
   a_2=nV_0+\frac{T}{\overline{c}\,(1-\overline{c})},\qquad  a_3=-\frac{(1-2\overline{c})}{2\,\overline{c}^{\,2}(1-\overline{c})^2}\,T,\qquad    %
   a_4=\frac{(1-3\overline{c}+3\overline{c}^{\,2})}{3\,\overline{c}^{\,3}(1-\overline{c})^3}\,T, %
}\vspace{2mm}\\ %
\displaystyle{%
   K=-\frac{2\pi}{3}\,n\!\int \! dr\,r^4V(r), \qquad %
   \varphi_0(T,\overline{c})=\frac{\overline{c}^{\,2}n}{2}\,V_0 + T\big[\,\overline{c}\ln \overline{c} + (1-\overline{c})\ln (1-\overline{c})\big], %
}%
\end{array}
\end{equation}
where $V_0=\!\int d{\bf r}V(r)$. The coefficient $a_2$ can change
sign at some temperature becoming negative, if only $V_0<0$, so that
the possibility of a loss of stability is determined by the nature
of the interaction between atoms [1]. The temperature, at which the
coefficient $a_2$ changes sign, is given by formula
\begin{equation} \label{11}
\begin{array}{l}
\displaystyle{%
   T_p=n|V_0|\,\overline{c}\,(1-\overline{c}).    %
}%
\end{array}
\end{equation}
The coefficient $a_2=a_0\tau$ includes the parameter
\begin{equation} \label{12}
\begin{array}{l}
\displaystyle{%
   a_0\equiv n|V_0|,   %
}%
\end{array}
\end{equation}
and the other two expansion coefficients have the form
\begin{equation} \label{13}
\begin{array}{ccc}
\displaystyle{%
   a_3=-\frac{(1-2\overline{c})}{2\,\overline{c}^{\,2}(1-\overline{c})^2}\,T,\qquad    %
   a_4=\frac{(1-3\overline{c}+3\overline{c}^{\,2})}{3\,\overline{c}^{\,3}(1-\overline{c})^3}\,T. %
}%
\end{array}
\end{equation}
The coefficient $K$ in this model does not depend on temperature. We
represent the coefficients (13) as $a_3\equiv -T q_3(\overline{c})$,
$a_4\equiv T q_4(\overline{c})$,where functions are introduced which
depend only on the average concentration:
\begin{equation} \label{14}
\begin{array}{ccc}
\displaystyle{%
   q_3(\overline{c})\equiv\frac{(1-2\overline{c})}{2\,\overline{c}^{\,2}(1-\overline{c})^2},\qquad    %
   q_4(\overline{c})\equiv\frac{(1-3\overline{c}+3\overline{c}^{\,2})}{3\,\overline{c}^{\,3}(1-\overline{c})^3}. %
}%
\end{array}
\end{equation}
In paper [1] the authors considered in detail the case when the
coefficient $a_3=0$. As we can see, in the considered approximation
of a homogeneous medium the coefficient at the cubic term in
expansion (7) vanishes only at the average concentration $\overline{c}=1/2$. %

\section{Transition to spatially inhomogeneous states}\vspace{-0mm} %
Let us consider the transitions of a binary solution to spatially
inhomogeneous states, assuming that the concentration can depend
only on one spatial coordinate $x$, which varies within the limits
$-L/2\leq x\leq L/2$. In this one-dimensional case, the total free
energy (1) can be written as
\begin{equation} \label{15}
\begin{array}{l}
\displaystyle{%
   \overline{F}=nA\!\int \!dx\left[\frac{K}{2}\left(\frac{d\,\delta c}{dx}\right)^2+a_2\frac{(\delta c)^2}{2}+a_3\frac{(\delta c)^3}{3}+a_4\frac{(\delta c)^4}{4}\right],   %
}
\end{array}
\end{equation}
where $A$ is the area of a sample, the normal to which is parallel
to the $x$ axis. Equation (8) in this case takes the form
\begin{equation} \label{16}
\begin{array}{l}
\displaystyle{%
   K\frac{d^2\delta c}{dx^2}=a_2\delta c+a_3(\delta c)^2+a_4(\delta c)^3.  %
}
\end{array}
\end{equation}
The first integral of this equation is
\begin{equation} \label{17}
\begin{array}{l}
\displaystyle{%
   \frac{K}{2}\!\left(\frac{d\,\delta c}{dx}\right)^2=C-U(\delta c),   %
}
\end{array}
\end{equation}
where $C$ is the constant of integration, and
\begin{equation} \label{18}
\begin{array}{l}
\displaystyle{%
   U(\delta c)=-a_4\frac{(\delta c)^4}{4}-a_3\frac{(\delta c)^3}{3}-a_2\frac{(\delta c)^2}{2}.    %
}
\end{array}
\end{equation}
Equation (17) is similar to the equation of motion of a material
point in an external field $U(\delta c)$ (18) with energy $C$.
Taking into account formulas (12)\,--\,(14), the field (18) can be
represented as $U(\delta c)\equiv T q_4\bar{U}(c)\big/4$:
\begin{equation} \label{19}
\begin{array}{ccc}
\displaystyle{%
   \bar{U}(c)=-\Big[(\delta c)^2-2q(\overline{c})(\delta c)+p(\overline{c})\frac{\tau}{1+\tau}\Big](\delta c)^2 =  %
   -(c-c_{01})(c-c_{02})(c-\overline{c})^2, %
}\vspace{2mm}\\ %
\end{array}
\end{equation}
where
\begin{equation} \label{20}
\begin{array}{ccc}
\displaystyle{%
   q(\overline{c})\equiv\frac{2}{3}\frac{q_3}{q_4}\equiv\frac{(1-2\overline{c})\,\overline{c}\,(1-\overline{c})}{(1-3\overline{c}+3\overline{c}^2)}, \qquad %
   p(\overline{c})=\frac{2a_0}{T_pq_4}\equiv\frac{6\overline{c}^2(1-\overline{c})^2}{(1-3\overline{c}+3\overline{c}^2)}.%
} %
\end{array}
\end{equation}
The concentrations at which the field vanishes are determined by the formulas %
\begin{equation} \label{21}
\begin{array}{ccc}
\displaystyle{%
   c_0=\overline{c}, \qquad c_{01}=\overline{c}+q-D, \qquad c_{02}=\overline{c}+Q+D,  %
} %
\end{array}
\end{equation}
where
\begin{equation} \label{22}
\begin{array}{ccc}
\displaystyle{%
   D\equiv D(\tau)\equiv \sqrt{\frac{p}{1+\tau_C}}\sqrt{\frac{\tau_C-\tau}{1+\tau}},  %
} %
\end{array}
\end{equation}
and temperature
\begin{equation} \label{23}
\begin{array}{ccc}
\displaystyle{%
   \tau_C=\frac{6(1-3\overline{c}+3\overline{c}^2)}{(5-14\overline{c}+14\overline{c}^2)}-1  %
} %
\end{array}
\end{equation}
is found from the condition $\big(q^2\big/p\big)-\tau_C\big/(1+\tau_C)=0$. %
Inhomogeneous phases can exist at temperatures $\tau<\tau_C$, when
the radical expression in (22) is positive. As was noted, for
$\overline{c}=1/2$ the cubic term in (18) vanishes, and the field
$U(\delta c)$ becomes symmetric with respect to the axis
$c=\overline{c}$.  In this case $\tau_C=0$. This case was in detail
considered by the authors earlier in [1]. At other concentrations
$q\neq 0$, the field $U(\delta c)$ turns out to be asymmetric, and
$\tau_C\neq 0$. The form of the function %
$\bar{U}(c)=-(c-c_{01})(c-c_{02})(c-\overline{c})^2$ (19) is shown
in Fig.\,1. Its maximums are located at the points
\begin{equation} \label{24}
\begin{array}{ccc}
\displaystyle{%
   c_{m1}=\overline{c}+\frac{3}{4}\Big(q-\frac{1}{2}G\Big), \qquad   %
   c_{m2}=\overline{c}+\frac{3}{4}\Big(q+\frac{1}{2}G\Big),          %
} %
\end{array}
\end{equation}
where $G\equiv\frac{2}{3}\sqrt{q^2+8D^2}$.  The field values in maximums: %
\begin{equation} \label{25}
\begin{array}{ccc}
\displaystyle{%
   \left.
     \begin{array}{l}
       \bar{U}(c_{m1}) \\
       \bar{U}(c_{m2})
     \end{array} \!\right\}
   =\frac{9}{16}\bigg[D^2-\frac{1}{16}\bigg(q\pm\frac{3}{2}G\bigg)^{\!2}\bigg]\!\bigg(q\mp\frac{1}{2}G\bigg)^{\!2}. %
}%
\end{array}
\end{equation}
In the symmetric case (Fig.\,1\textit{a}) at $\overline{c}=1/2$,
when $q=0$, we have $\bar{U}(c_{m1})=\bar{U}(c_{m2})=D^2\!\big/4$.
\vspace{0mm} %
\begin{figure}[h!]
\vspace{-0mm}  \hspace{0mm}
\includegraphics[width = 15cm]{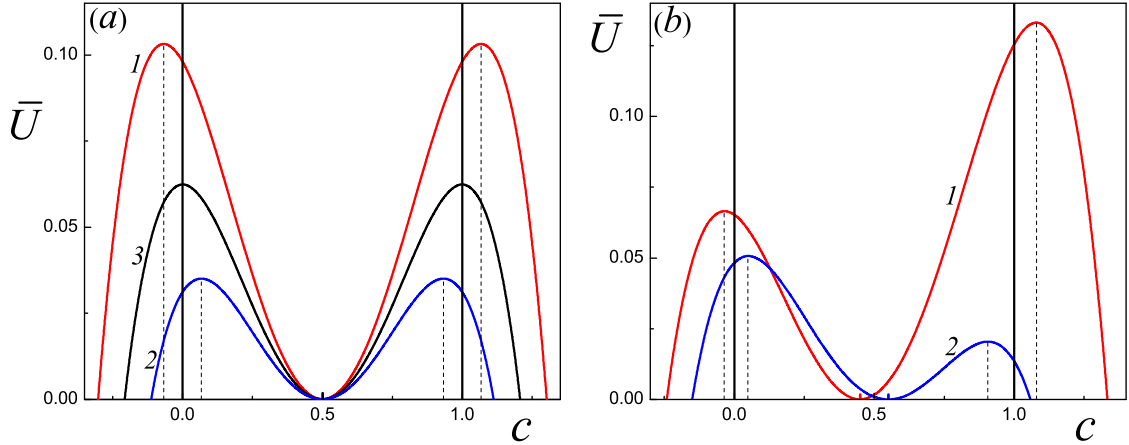} 
\vspace{-4mm} %
\caption{\label{fig01} 
Form of the field $\bar{U}(c)$ (19): %
(\!{\it a}) the symmetric case $\overline{c}=1/2$ at temperatures: %
({\it 1}) $\tau=-0.30$, ({\it 2}) $\tau=-0.20$, \newline ({\it 3}) $\tau=-0.25$; %
(\!{\it b}) the asymmetric case: %
({\it 1}) $\overline{c}=0.45, \,\tau=-0.30$, ({\it 2}) $\overline{c}=0.55, \,\tau=-0.20$. %
}%
\end{figure}

There are two characteristic temperature regions, where the states
are qualitatively different. In region I the values of
concentration, at which the field (19) reaches its maximum, lie in
the region of permissible values $0\leq c_{m1} < c_{m2} \leq 1$
(Fig.\,1, curves {\it 2}). In region II the maximums lie outside the
region of permissible values, so that the inequalities %
$c_{m1} < 0$ and $c_{m2} > 1$ are satisfied (Fig.\,1, curves {\it 1}). %
The boundary temperature $T_1$ is determined by the conditions %
$c_{m1} = 0, \, c_{m2} = 1$ and $\overline{c}=1/2$, so that
\begin{equation} \label{26}
\begin{array}{ccc}
\displaystyle{%
   \frac{T_1}{T_p}=\frac{3}{4}.   %
} %
\end{array}
\end{equation}
Thus, two temperature regions should be considered, where the states
are qualitatively different: region I at $T_1<T<T_p$ , which was
called the {\it K}\,-phase in [1], and region II at $0<T<T_1$ (the
{\it W}\,-phase).

Equation (17) can be represented as
\begin{equation} \label{27}
\begin{array}{l}
\displaystyle{%
   \xi^2\!\left(\frac{dz}{dx}\right)^{\!2}=(z-z_1)(z-z_2)(z-z_3)(z-z_4).   %
}
\end{array}
\end{equation}
Here and below we use the notation $z\equiv\delta c = c-\overline{c}$. %
In (27) $z_1\leq z_2\leq z_3\leq z_4$ are the real, arranged in
ascending order roots of the algebraic equation
\begin{equation} \label{28}
\begin{array}{l}
\displaystyle{%
   (z-q+D)(z-q-D)z^2+\varepsilon=0,   %
}
\end{array}
\end{equation}
where $\varepsilon\equiv 4C\big/Tq_4$, and the correlation length
$\xi$ in (27), which determines the size of the inhomogeneity
region, is given by the relation
\begin{equation} \label{29}
\begin{array}{l}
\displaystyle{%
   \xi^2\equiv\frac{2K}{Tq_4}.   %
}
\end{array}
\end{equation}

If all real roots of equation (28) are different, then equation (27)
has a periodic solution [9]
\begin{equation} \label{30}
\begin{array}{l}
\displaystyle{%
   z(x)=z_1+\frac{r(z_2-z_1)}{r-{\rm sn}^2\Big(\displaystyle{s\,\frac{x}{\xi}},m\Big)},   %
}
\end{array}
\end{equation}
where ${\rm sn}\Big(\displaystyle{s\,\frac{x}{\xi}},m\Big)$ is the
elliptic sine [10], and also
\begin{equation} \label{31}
\begin{array}{l}
\displaystyle{%
   s\equiv\frac{1}{2}\sqrt{(z_4-z_2)(z_3-z_1)}, \qquad %
   r\equiv\frac{(z_3-z_1)}{(z_3-z_2)}, \qquad          %
   m=\frac{(z_3-z_2)(z_4-z_1)}{(z_4-z_2)(z_3-z_1)}.   %
}%
\end{array}
\end{equation}
The period of the solution (30) is
\begin{equation} \label{32}
\begin{array}{l}
\displaystyle{%
   L_0=\frac{2K(m)}{s}\,\xi,   %
}
\end{array}
\end{equation}
where $K(m)$ the complete elliptic integral of the first kind [10].

\section{The case of concentrations other than one-half. The matching
condition}\vspace{-0mm} %
Let us first consider the case, when the average concentration is
different from $\overline{c}=1/2$. In this case the cubic term in
the free energy expansion (15) is nonzero. In this case, there exist
periodic solutions in the form of concentration waves (30). In order
for such solutions to describe real physical states, they must
satisfy the matching condition (3), which follows from the
requirement of conservation of the number of particles. In the
one-dimensional case under consideration this condition has the form
\begin{equation} \label{33}
\begin{array}{l}
\displaystyle{%
   \int_0^{L_0}\!z(x)dx=0,   %
}%
\end{array}
\end{equation}
where $L_0$ is the period (32).  Using equation (27), the matching
condition (33) can be represented as
\begin{equation} \label{34}
\begin{array}{l}
\displaystyle{%
   J\big(\alpha,\beta\big)=J\big(\beta,\alpha\big),  
}%
\end{array}
\end{equation}
where
\begin{equation} \label{35}
\begin{array}{l}
\displaystyle{%
   J\big(\alpha,\beta\big)\equiv\int_0^1\!  %
   \frac{\lambda\,d\lambda}{\sqrt{\big(\cos^{-1}\!\alpha-\lambda\big)\big(1-\lambda\big)\big(\psi(\alpha,\beta)+\lambda\big)\big(\cos^{-1}\!\beta\,\cdot\psi(\alpha,\beta)+\lambda\big)}}, %
}%
\end{array}
\end{equation}
$\displaystyle{\psi(\alpha,\beta)=\frac{1+\cos\beta}{1+\cos\alpha}}$, %
and the angles $\alpha, \beta$  are defined through the ratios of
the moduli of the roots of equation (28)
\begin{equation} \label{36}
\begin{array}{l}
\displaystyle{%
   \frac{|z_2|}{|z_1|}=\cos\alpha, \qquad \frac{|z_3|}{|z_4|}=\cos\beta,   %
}%
\end{array}
\end{equation}
so that $0<\alpha,\,\beta<\pi/2$. An analysis shows that the
condition (34) is satisfied in the single case $\alpha=\beta$, which
corresponds to the symmetric case with concentration $\overline{c}=1/2$. %
For all other concentrations the condition (34) is not fulfilled
and, consequently, such solutions are not realized.

In addition, aside from periodic solutions, when the condition
$\varepsilon=\bar{U}(z_{m1})$ is fulfilled, where $z_{m1}=c_{m1}-\overline{c}$, %
Eq.\,(27) has a solution localized near an arbitrary point $x_0$,
which can be represented as
\begin{equation} \label{37}
\begin{array}{l}
\displaystyle{%
   \frac{1}{\sqrt{\big(a_--a_m\big)\big(1-a_m\big)}}\,   %
   \ln\left[\frac{\left[\sqrt{\big(1-a_m\big)\big(a_--\lambda\big)}-\sqrt{\big(a_--a_m\big)\big(1-\lambda\big)}\,\right]^{\!2}}{|\lambda-a_m|}\right]= %
   \pm z_+\frac{\big(x-x_0\big)}{\xi},
}%
\end{array}
\end{equation}
where $\lambda\equiv z\big/z_+$, $a_m\equiv z_{m1}\big/z_+$, $a_-\equiv z_-\big/z_+$, and here %
\begin{equation} \label{38}
\begin{array}{l}
\displaystyle{%
   z_\pm=\frac{1}{4}\left(q+\frac{3}{2}\,G\pm 2\sqrt{q^2+\frac{3}{2}\,qG}\right),  \qquad   %
   z_{m1}=\frac{3}{8}\big(2q-G\big),
}%
\end{array}
\end{equation}
and as before $G\equiv\frac{2}{3}\sqrt{q^2+8D^2}$. This solution
satisfies the matching condition (33) only in the limit of a sample
of infinite length. Thus, at concentrations $\overline{c}\neq 1/2$, %
when the cubic term in the free energy is nonzero, the existing
spatially inhomogeneous periodic and localized solutions do not
satisfy the matching condition, and therefore cannot be physically
realized.

\section{Thermodynamics of inhomogeneous states at $\overline{c}=1/2$}\vspace{-0mm} %
At the average concentration of $\overline{c}=1/2$ the cubic term in
expansion (18) vanishes, since $q(1/2)=0$, and also in this case %
$q_4(1/2)=16/3$, $p(1/2)=3/2$. This case was considered in detail in
the authors' work [1]. Here, using a slightly different method, we
present the main results and introduce some refinements into the
method of describing the thermodynamics of inhomogeneous states.

In the temperature range $T_1<T<T_p$  ({\it K}\,-phase) at %
$\varepsilon=D^4\big/4$ there exist solutions in the form of a ``kink'' %
\begin{equation} \label{39}
\begin{array}{l}
\displaystyle{%
  z(x)=c(x)-\frac{1}{2}=\pm\frac{D}{\sqrt{2}}\,{\rm th}\bigg(\frac{D}{\sqrt{2}}\frac{x}{\xi}\bigg),  %
}%
\end{array}
\end{equation}
describing the stratification of the solution into regions with
enriched and depleted concentrations, as well as, at $0<\varepsilon<D^4\big/4$, %
there exist solutions in the form of concentration waves
\begin{equation} \label{40}
\begin{array}{l}
\displaystyle{%
  z(x)=c(x)-\frac{1}{2}=\pm\frac{D_-}{\sqrt{2}}\,{\rm sn}\bigg(\frac{D_+}{\sqrt{2}}\frac{x}{\xi};m\bigg),  %
}%
\end{array}
\end{equation}
where $m=D_-^2\big/D_+^2$, $D_\pm\equiv\sqrt{D^2\pm\sqrt{D^4-4\varepsilon}}$. %
In the temperature range $0<T<T_1$ ({\it W}\,-phase), there are
physical solutions only in the form of concentration waves (40).
Solutions (39), (40) satisfy the matching conditions (3), (33).
Since $\varepsilon=D^4m\big/(1+m)^2$, $D_+^2=2D^2\big/(1+m)$, $D_-^2=2D^2m\big/(1+m)$, %
then each of the solutions is characterized by a parameter $m$
numerating all possible inhomogeneous states. For $m=1$, solution
(40) transforms into the ``kink'' (39). The temperature dependence
of quantities is mainly determined by the function (22), which in
this case takes the form
\begin{equation} \label{41}
\begin{array}{l}
\displaystyle{%
  D^2\equiv  D^2(\tau)=-\frac{3}{2}\frac{\tau}{(1+\tau)}.  %
}%
\end{array}
\end{equation}
The width of the transition region in the ``kink'' $L_K(\tau)$ and
the period of the concentration wave $L_0(m;\tau)$ are determined by the formulas %
\begin{equation} \label{42}
\begin{array}{l}
\displaystyle{%
  L_K\big(T\big)=\sqrt{2}\frac{\xi_0}{\sqrt{1-T/T_p}},  \qquad %
  L_0\big(m;T\big)=4K(m)\sqrt{1+m}\frac{\xi_0}{\sqrt{1-T/T_p}},
}%
\end{array}
\end{equation}
where $\xi_0^2=K/a_0$ is the temperature-independent correlation
length, and $\xi^2=3\xi_0^2\big/2(1+\tau)$ (29). The temperature
dependences of the lengths (42) are shown in Fig.\,2. When
approaching the homogeneous phase $T\rightarrow T_p$, they tend to
infinity, and as $T\rightarrow 0$ they tend to a finite value $\sim \xi_0$. %
\vspace{0mm} %
\begin{figure}[h!]
\vspace{-0mm}  \hspace{0mm}
\includegraphics[width = 7.9cm]{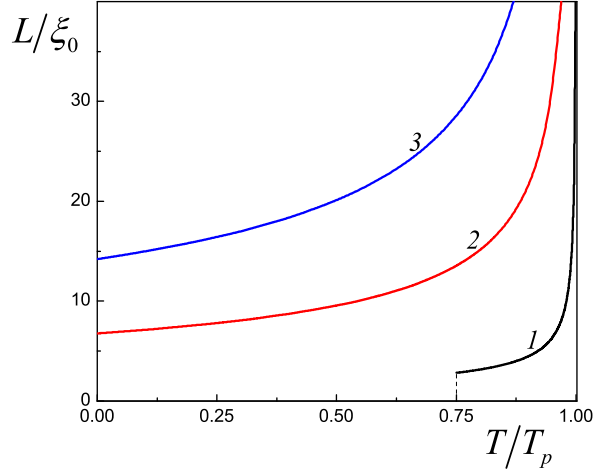} 
\vspace{-4mm} %
\caption{\label{fig02} 
Temperature dependencies of the inhomogeneity dimensions: %
({\it 1}) ``kink'' width $L_K\big(T\big)$; %
wave period $L_0\big(m;T\big)$  for \newline ({\it 2}) $m=0.1$ and ({\it 3}) $m=0.9$. %
}%
\end{figure}

Let us calculate the contribution of inhomogeneities to the
thermodynamic quantities of the solution at the concentration $\overline{c}=1/2$. %
According to (5),\,(6), the entropy and energy can be represented in
the form $S=-N\big(\partial\varphi_0/\partial T\big)+\delta S$, %
$E=N\big(\varphi_0-T\partial\varphi_0/\partial T\big)+\delta E$. %
We will be interested in the contributions of inhomogeneities to
these quantities, defined by the formulas
\begin{equation} \label{43}
\begin{array}{l}
\displaystyle{%
  \delta S = -\frac{4}{3}nA\Big(\frac{3}{2}J_2+J_4\Big), %
}%
\end{array}
\end{equation}
\vspace{-4mm}%
\begin{equation} \label{44}
\begin{array}{l}
\displaystyle{%
  \delta E = 2nA T_p\big(\xi_0^2J - J_2\big). %
}%
\end{array}
\end{equation}
Here we introduce the notation of the following integrals:
\begin{equation} \label{45}
\begin{array}{l}
\displaystyle{%
  J\equiv\int\!\bigg(\frac{dz}{dx}\bigg)^{\!2}dx, \qquad  %
  J_2\equiv\int\!z^2dx, \qquad J_4\equiv\int\!z^4dx.  %
}%
\end{array}
\end{equation}
For the ``kink'' (39) in a sample of length $L$, these integrals are as follows: %
\begin{equation} \label{46}
\begin{array}{l}
\displaystyle{%
  J=\frac{D^3}{\sqrt{2}\,\xi}\Big({\rm th}y - \frac{1}{3}\,{\rm th}^3y\Big),  \qquad %
  J_2=\sqrt{2}D\xi\big(y-{\rm th}y \big),  \qquad %
  J_4=\frac{D^3}{\sqrt{2}}\,\xi\Big(y - {\rm th}y - \frac{1}{3}\,{\rm th}^3y\Big),  %
}%
\end{array}
\end{equation}
where $y\equiv \big(D\big/2\sqrt{2}\big)\big(L/\xi\big)$. For the
concentration waves, integrals (45) are expressed by the formulas
\begin{equation} \label{47}
\begin{array}{l}
\displaystyle{%
  J=\frac{L}{\xi^2}D^4\frac{m}{(1+m)^2}\frac{I_0(m)}{K(m)},  \qquad %
  J_2=LD^2\frac{m}{(1+m)}\frac{I_2(m)}{K(m)},  \qquad %
  J_4=LD^4\frac{m^2}{(1+m)^2}\frac{I_4(m)}{K(m)}. %
}%
\end{array}
\end{equation}
Here
\begin{equation} \label{48}
\begin{array}{ccc}
\displaystyle{%
  I_0(m)=\frac{1}{3m}\big[(m-1)K(m)+(m+1)E(m)\big],  %
}\vspace{2mm}\\ %
\displaystyle{\hspace{0mm}%
  I_2(m)=\frac{1}{m}\big[K(m)+E(m)\big],  %
}\vspace{2mm}\\ %
\displaystyle{\hspace{0mm}%
  I_4(m)=\frac{1}{3m^2}\big[(m+2)K(m)-2(m+1)E(m)\big],  %
}%
\end{array}
\end{equation}
where $K(m),\,E(m)$ are the complete elliptic integrals of the first
and second kind [10].

For the entropy and energy per a particle in the ``kink'', for $L\gg
\xi$, we get:
\begin{equation} \label{49}
\begin{array}{l}
\displaystyle{%
  s_K\equiv\frac{\delta S_K}{N}=-D^2-\frac{1}{3}D^4, %
}%
\end{array}
\end{equation}
\vspace{-4mm}
\begin{equation} \label{50}
\begin{array}{l}
\displaystyle{%
  e_K\equiv\frac{\delta E_K}{N}=-T_pD^2. %
}%
\end{array}
\end{equation} \newpage
The contribution of the ``kink'' to the free energy:
\begin{equation} \label{51}
\begin{array}{l}
\displaystyle{%
  \psi_K\equiv \frac{\delta F_K}{N}=e_K-Ts_K=(T-T_p)D^2+\frac{T}{3}D^4. %
}%
\end{array}
\end{equation}

Taking into account formulas (47),\,(48), for the entropy and energy
of the concentration wave per a particle we obtain:
\begin{equation} \label{52}
\begin{array}{ccc}
\displaystyle{%
  s(m)\equiv\frac{\delta S_W(m)}{N}=-2\Big[g(m)D^2+\frac{2}{9}f(m)D^4\Big],  %
}%
\end{array}
\end{equation}
\vspace{-4mm}
\begin{equation} \label{53}
\begin{array}{ccc}
\displaystyle{%
  e(m)\equiv\frac{\delta E_W(m)}{N}=T_p\Big[\!-2g(m)D^2+\frac{4}{9}(1+\tau)\big(3g(m)-2f(m)\big)D^4\Big].  %
}%
\end{array}
\end{equation}
Here the functions are introduced
\begin{equation} \label{54}
\begin{array}{ccc}
\displaystyle{%
  f(m)\equiv\frac{(2+m)}{(1+m)^2}-\frac{2}{(1+m)}\frac{E(m)}{K(m)} %
}%
\end{array}
\end{equation}
and
\begin{equation} \label{55}
\begin{array}{ccc}
\displaystyle{%
  g(m)=\frac{1}{(1+m)}\bigg(1-\frac{E(m)}{K(m)}\bigg).   %
}%
\end{array}
\end{equation}
There is a relationship between these functions $f(m)=-m\big/(1+m)^2 + 2g(m)$. %
The form of functions (54),\,(55) is shown in Fig.\,3.
\vspace{0mm} %
\begin{figure}[h!]
\vspace{-0mm}  \hspace{0mm}
\includegraphics[width = 7.3cm]{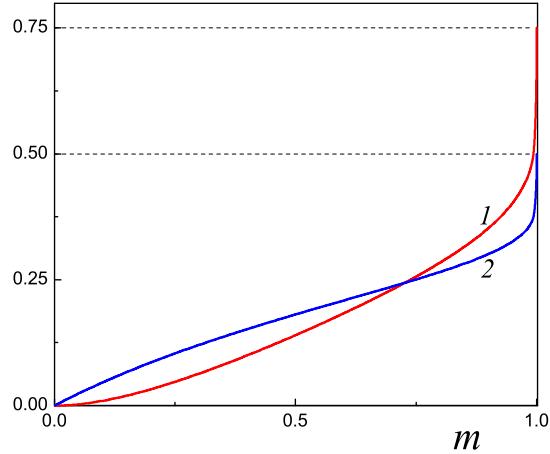} 
\vspace{-3mm} %
\caption{\label{fig03} 
Functions: ({\it 1}) $f(m)$\,\,(54), ({\it 2}) $g(m)$\,\,(55). %
}%
\end{figure}

The free energy per a particle:
\begin{equation} \label{56}
\begin{array}{ccc}
\displaystyle{%
  \psi(m)\equiv\frac{\delta F_W(m)}{N}=e(m)-Ts(m)= %
  2\,T_pg(m)\tau D^2 + \frac{4}{9}T_p(1+\tau)\big(3g(m)-f(m)\big)D^4.  %
}%
\end{array}
\end{equation}
The free energy (56) can be expressed in terms of  one function (54): %
\begin{equation} \label{57}
\begin{array}{ccc}
\displaystyle{%
  \psi(m)=-T_p\frac{\tau^2}{(1+\tau)}f(m).  %
}%
\end{array}
\end{equation}
Note that near the temperature of transition to the inhomogeneous
state, the entropy (49),\,(52) and the energy (50),\,(53) are linear
in $\tau$, while the free energy (51),\,(56),\,(57) is quadratic.
Taking into account that $f(1)=3/4$ and $g(1)=1/2$, formulas %
(52),\,(53),\,(56) transform into formulas (49),\,(50),\,(51) for
the ``kink''.

Formulas (49)\,--\,(57) determine the contribution of a specific
solution of the nonlinear equation (27) to the entropy, energy and
free energy. When finding the total thermodynamic quantities, the
contributions of all solutions should be taken into account.  The
contribution of the solution will be the more significant, the lower
the free energy associated with it. According to the general
principles of statistical physics, the probability $dw=w(m)dm$ %
of finding a system in a state with parameter $m$ in the interval %
$m\div m+dm$ is determined by a ``single-particle'' distribution function over states $m$: %
$\displaystyle{w(m;\tau)\sim\exp\!\bigg(\!-\frac{\psi(m;\tau)}{T}\bigg)=\exp\!\bigg(\frac{\tau^2}{(1+\tau)^2}f(m)\bigg)}$. %
Function $f(m)$ is increasing and reaches its maximum value
$f(1)=3/4$ at $m=1$ (Fig.\,3, curve {\it 1}). With a small deviation
from $m=1$, it decreases very quickly from its maximum value as %
$f(m)\approx f(1)-2\big/\ln\!\!\big[16/(1-m)\big]$. Therefore, it is
more convenient to define the distribution function in terms of the
difference $\Delta f(m)\equiv f(m)-f(1)$, so that we set
\begin{equation} \label{58}
\begin{array}{ccc}
\displaystyle{%
  w(m;\tau)=\frac{1}{Z(\tau)}\exp\!\bigg(\frac{\tau^2}{(1+\tau)^2}\Delta f(m)\bigg). %
}%
\end{array}
\end{equation}
The statistical integral $Z(\tau)$ is defined by the normalization
condition $\int_0^{m_*(\tau)}\!w(m;\tau)dm=1$:
\begin{equation} \label{59}
\begin{array}{ccc}
\displaystyle{%
  Z(\tau)=\int_0^{m_*(\tau)}\!e^{\frac{\tau^2}{(1+\tau)^2}\Delta f(m)}dm, %
}%
\end{array}
\end{equation}
where the integration is carried out over all permissible values of
the parameter $m$, which vary from zero to the maximum value $m_*$.
Since $f(m)$ is an increasing function, then the solutions close to
the maximum value  $m_*$ make the main contribution to the
thermodynamic quantities. In the {\it K}\,-phase the solution with
the maximum $m_*=1$ is the ``kink'' (39). Taking into account that
${\rm sn}u\approx{\rm th}u$ [10] at $m\rightarrow 1$, in this limit
solution (40) goes over to (39). Thus, in the {\it K}\,-phase, where
$m_*=1$, the main contribution to the thermodynamic quantities is
made by the solution in the form of the ``kink''. In the {\it
W}\,-phase, the maximum value of the parameter $m$ depends on temperature %
\begin{equation} \label{60}
\begin{array}{ccc}
\displaystyle{%
  m_*(\tau)=\frac{1}{\big(4D^2-1\big)}=-\frac{(1+\tau)}{(1+7\tau)}=\frac{T/T_p}{6-7\big(T/T_p\big)}, %
}%
\end{array}
\end{equation}
and thermodynamics is determined only by solutions of the form of
concentration waves. The temperature dependence of the parameter
$m_*(\tau)$ (60) is shown in Fig.\,4. \newline
\vspace{-3mm} %
\begin{figure}[h!]
\vspace{-0mm}  \hspace{0mm}
\includegraphics[width = 7.6cm]{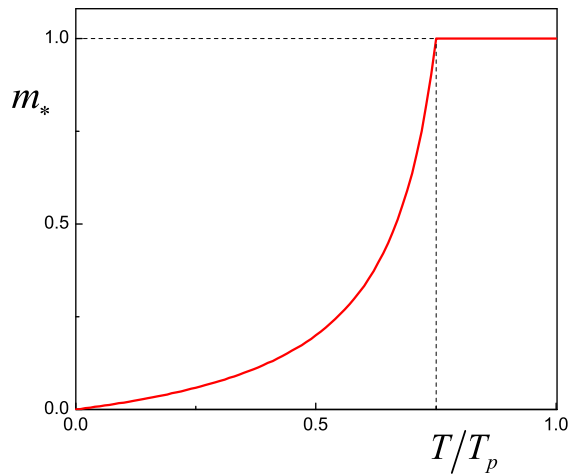} 
\vspace{-3mm} %
\caption{\label{fig04} 
Dependence of the maximum value of the parameter $m\equiv m_*(T)$ on
temperature.
}%
\end{figure}

\noindent The form of distribution functions in both inhomogeneous
phases is shown in Fig.\,5.
\vspace{0mm} %
\begin{figure}[h!]
\vspace{-0mm}  \hspace{0mm}
\includegraphics[width = 14.9cm]{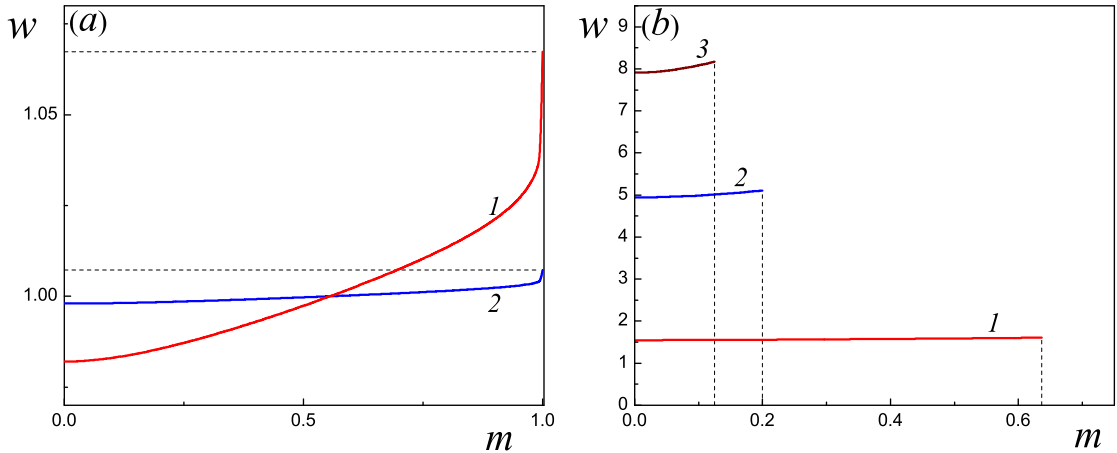} 
\vspace{-3mm} %
\caption{\label{fig05} 
Distribution functions $w(m;\tau)$ at certain temperatures: %
({\it a}) in the {\it K}\,-phase at ({\it 1}) $\tau=-0.25$, ({\it 2}) $\tau=-0.1$; %
({\it b}) in the {\it W}\,-phase at ({\it 1}) $\tau=-0.3$, ({\it 2}) $\tau=-0.5$, ({\it 3}) $\tau=-0.6$. %
}%
\end{figure}

The average value of an arbitrary function $\Phi(m)$ is found by the formula %
\begin{equation} \label{61}
\begin{array}{ccc}
\displaystyle{%
  \big\langle\Phi\big\rangle = \int_0^{m_*}\!\Phi(m)\,w(m;\tau)\,dm. %
}%
\end{array}
\end{equation}
Averaging expressions (52),\,(53) and (57), we obtain the average
values of entropy, energy and free energy, which are expressed
through the average values of functions (54),\,(55):
\begin{equation} \label{62}
\begin{array}{ccc}
\displaystyle{%
  \langle s\rangle = -2\Big[ \langle g\rangle D^2 + \frac{2}{9}\langle f\rangle D^4 \Big], %
}%
\end{array}
\end{equation}
\vspace{-3mm}
\begin{equation} \label{63}
\begin{array}{ccc}
\displaystyle{%
  \langle e\rangle =T_p\Big[ -2\langle g\rangle D^2 + \frac{4}{9}(1+\tau)\big(3\langle g\rangle - 2\langle f\rangle \big)D^4 \Big], %
}%
\end{array}
\end{equation}
\vspace{-3mm}
\begin{equation} \label{64}
\begin{array}{ccc}
\displaystyle{%
  \langle \psi\rangle = -T_p\frac{\tau^2}{(1+\tau)}\langle f\rangle .  %
}%
\end{array}
\end{equation}
\newpage\noindent
The heat capacity per a particle at constant volume follows from
formula (62) for the entropy:
\begin{equation} \label{65}
\begin{array}{ccc}
\displaystyle{%
  \frac{C_V}{N} = T \frac{d\langle s\rangle }{dT} = %
  -\bigg\{2D^2\langle g\rangle\bigg[T\frac{d\ln\langle g\rangle}{dT}+\frac{1}{\tau}\bigg]+ %
  \frac{4}{9}D^4\langle f\rangle\bigg[T\frac{d\ln\langle f\rangle}{dT}+\frac{2}{\tau}\bigg]\bigg\}. %
}%
\end{array}
\end{equation}
From formulas (62)\,--\,(64) it also follows that there holds the
identity $\langle \psi\rangle = \langle e\rangle - T\langle s\rangle $. %
Note that in work [1] the entropy was defined through the average of
the logarithm of the one-particle distribution function, as is
customary in the theory of rarefied gases [11]. However, in this
case the direct calculation of the entropy, which is based on the
use of formula (9) for the free energy, shows that the entropy of
the inhomogeneous state is not reduced to the average of the
logarithm of the distribution function. Therefore, in respect to the
calculation of the entropy and heat capacity, we corrected the
results of work [1]. The results of a new calculation of the
temperature dependences of the entropy and heat capacity by the
formulas (62),\,(65) are shown in Fig.\,6.
\vspace{0mm} %
\begin{figure}[h!]
\vspace{-1mm}  \hspace{0mm}
\includegraphics[width = 15.3cm]{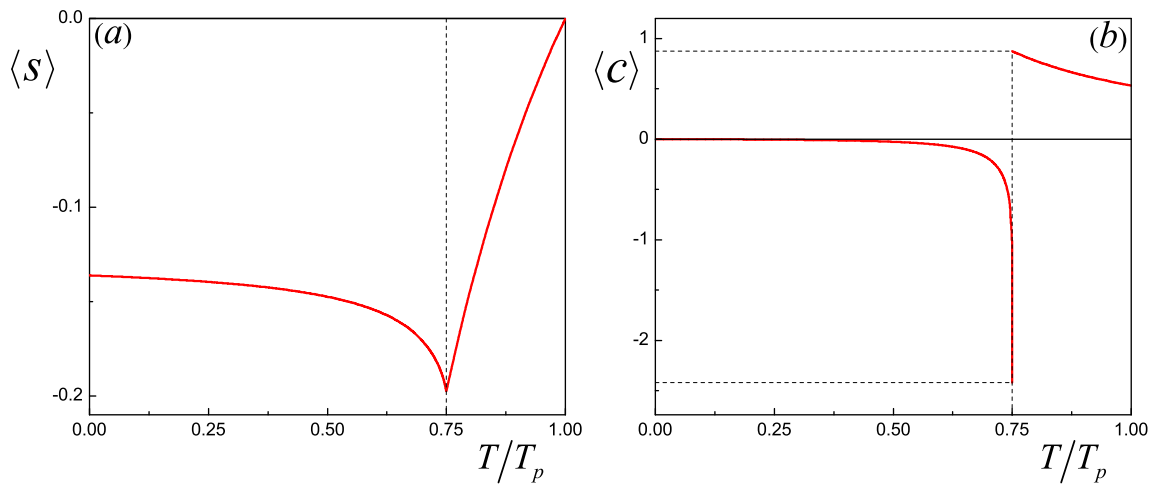} 
\vspace{-3mm} %
\caption{\label{fig06} 
Temperature dependencies of the inhomogeneity contribution: %
({\it a}) to entropy, ({\it b}) to heat capacity.
}%
\end{figure}
\vspace{-6mm} %

\section{Transitions between phases. Heat capacity.}\vspace{-3mm} %
Above the temperature $T_p$ the solution is in the homogeneous state
({\it H}\,-phase). In the temperature range $T_1<T<T_p$, there
arises the {\it K}\,-phase in which the states both in the form of
the ``kink'' (39) and in the form of the concentration waves (40)
are possible. For the ``kink'' the parameter $m\equiv m_*=1$, %
and for the waves in this phase it varies in the interval $0\leq m <1$, %
but the main contribution to thermodynamics is made by the ``kink''.
Consider the transition from the homogeneous state to the {\it
K}\,-phase. Near the transition temperature at $|\tau|\ll 1$, the
entropy (62) is $\langle s\rangle\approx 3\langle g\rangle_0\tau$,
where $\langle g\rangle_0=\int_0^1\!g(m)dm$. Therefore, during the
transition from the {\it H}\,-phase to the {\it K}\,-phase the heat
capacity undergoes a jump
\begin{equation} \label{66}
\begin{array}{ccc}
\displaystyle{%
  \frac{\Delta C}{N} = 3\langle g\rangle_0\approx 0.533.
}%
\end{array}
\end{equation}
Note that the jump in the isochoric heat capacity per a particle
(66) is a number which does not depend on the parameters of the
system and thermodynamic quantities.

In the temperature range $0<T<T_1$ , there is the {\it W}\,-phase in
which there are possible states only in the form of the
concentration waves (40). In this phase the parameter changes in the
interval $0\leq m \leq m_*$, where $m_*$ is defined by formula (60)
(Fig.\,4). With a decrease in temperature at $T_1\big/T_p=3/4$ or $\tau_1=-1/4$, %
a phase transition from the {\it K}\,- to the {\it W}\,-phase
occurs. Near the temperature of transition between these
inhomogeneous phases the entropy can be represented as
\begin{equation} \label{67}
\begin{array}{ccc}
\displaystyle{%
  \langle s\rangle = \langle s\rangle_1 + \zeta_1\Delta\tau +
  \zeta_2 \Delta m_*,
}%
\end{array}
\end{equation}
where $\Delta\tau=\tau-\tau_1$,
\begin{equation} \label{68}
\begin{array}{ccc}
\displaystyle{%
  \Delta m_*(\tau)=1-m_*(\tau)=
   \left\{
     \begin{array}{l}
       \,\,\,\, 0, \qquad \Delta\tau>0, \vspace{2mm} \\
       \displaystyle{\frac{32}{3}}\Delta\tau, \,\,\,\, \Delta\tau<0,
     \end{array} \!\right.
}%
\end{array}
\end{equation}
and also
\begin{equation} \label{69}
\begin{array}{ccc}
\displaystyle{%
  \langle s\rangle_1 = -\Big(\frac{7}{12} + \langle \Delta g\rangle_1 + \frac{1}{9}\langle \Delta f\rangle_1\Big), %
}%
\end{array}
\end{equation}
\vspace{-4mm}
\begin{equation} \label{70}
\begin{array}{ccc}
\displaystyle{%
  \zeta_1 = \frac{32}{27}\Big[ \Big(\big\langle\Delta f\Delta g\big\rangle_1-\langle\Delta f\rangle_1\langle\Delta g\rangle_1\Big) + %
  \frac{1}{9}\Big( \big\langle(\Delta f)^2\big\rangle_1 - \langle\Delta f\rangle_1^2 \Big)\Big] + %
  \frac{16}{3}\Big[\frac{2}{3} + \langle\Delta g\rangle_1 + \frac{2}{9}\langle\Delta f\rangle_1\Big], %
}%
\end{array}
\end{equation}
\vspace{-4mm}
\begin{equation} \label{71}
\begin{array}{ccc}
\displaystyle{%
  \zeta_2 = w(1;\tau_1)\Big(\langle \Delta g\rangle_1 + \frac{1}{9}\langle \Delta f\rangle_1\Big). %
}%
\end{array}
\end{equation}
Here $\langle\ldots\rangle_1$  means averaging with the distribution
function $w(m;\tau_1)$, and $w(1;\tau_1)=1/Z(\tau_1)$. From formulas
(67)\,--\,(71) it follows that during the transition between the
inhomogeneous phases the heat capacity undergoes a negative jump (Fig.\,6{\it b}) %
\begin{equation} \label{72}
\begin{array}{ccc}
\displaystyle{%
  \frac{\Delta C}{N}=\frac{\big(C_W-C_K\big)}{N} = 8\zeta_2. %
}%
\end{array}
\end{equation}

The qualitative difference between the results of this calculation
in comparison with the previous one [1] is that the transition from
the homogeneous state to the {\it K}\,-phase is the second-order
phase transition with a jump in the heat capacity. The transition
from the high-temperature inhomogeneous {\it K}\,-phase to the
low-temperature inhomogeneous {\it W}\,-phase remains, just as in
[1], the second-order transition, but with a negative jump in the
heat capacity (Fig.\,6{\it b}).

Let us also consider the behavior of the entropy and heat capacity
in the low-temperature limit $T\rightarrow 0$, when
$1+\tau\equiv\alpha\ll 1$. Since in this case $m_*\ll 1$, then there
hold the approximations $f(m)\approx 9m^2/8$, $g(m)\approx 0.5\big(m-7m^2/8\big)$. %
In view of this we find
\begin{equation} \label{73}
\begin{array}{ccc}
\displaystyle{%
  \langle s\rangle = s_0 - s_0'\frac{T}{T_p}, %
}%
\end{array}
\end{equation}
\vspace{-4mm}
\begin{equation} \label{74}
\begin{array}{ccc}
\displaystyle{%
  \frac{C}{N} = - s_0'\frac{T}{T_p}, %
}%
\end{array}
\end{equation}
where
\begin{equation} \label{75}
\begin{array}{ccc}
\displaystyle{%
  s_0=-\sqrt{2}\rho_1-\rho_2, \qquad %
  s_0'=\frac{\sqrt{2}\,e^{1\!/32}}{192\,r_0(0)}\Big[\big(1-4\sqrt{2}\rho_1\big)+\Big(\frac{1}{8}-4\rho_2\Big)\Big], %
}%
\end{array}
\end{equation}
and here $\rho_n=r_n(0)/r_0(0)$, $r_n(0)=\int_0^{\sqrt{2}/8}e^{y^2}y^n\,dy$. %
Thus, the contribution of inhomogeneity effects to the heat capacity
at $T\rightarrow 0$ tends to zero in proportion to temperature, and
the entropy tends to the constant value $s_0$ linearly with
temperature. Note that this does not contradict the basic principles
of thermodynamics, since under the used classical description the
entropy is determined up to a constant value, and a physical meaning
has a difference of entropies. To estimate the temperature range in
which the influence of quantum effects can be neglected, we use the
requirement of smallness of the thermal de Broglie wavelength
$\Lambda_B\sim\hbar\big/\sqrt{MT}$ ($M$ is the reduced mass of
particles in a solution) in comparison with the period of the
concentration wave. Since the wave period tends to the correlation
length $\xi_0=\sqrt{K/a_0}$ in the low-temperature limit, we have
the condition $\xi_0\gg \Lambda_B$. From it there follows the
limitation on temperature values at which the classical description is permissible: %
\begin{equation} \label{76}
\begin{array}{ccc}
\displaystyle{%
  T\gg\frac{\hbar^2}{M\xi_0^2}. %
}%
\end{array}
\end{equation}
For $M\sim 10^{-22}$\,g and $\xi_0\sim 10^{-8}$\,cm, it gives $T\gg 1$\,K. %

\section{Conclusion }\vspace{-2mm} %
In this work on the basis of a theoretical approach to the
description of a binary solution, previously proposed by the authors
in [1], the role of the cubic in concentration term in the expansion
of the free energy is studied and a refined calculation of the
contribution of inhomogeneity to thermodynamic quantities is given.
It is shown that taking into account the cubic term contributes to
the stability of the spatially homogeneous state of the solution. In
the considered model of an isotropic medium, the phase transition to
the inhomogeneous state proves to be possible only at half the
concentration. At other concentrations solutions of the equation for
the concentration in the form of concentration waves also exist, but
they do not satisfy the matching condition which follows from the
requirement of conservation of the number of particles, and
therefore are not realized. Note, however, that our calculations
were carried out under the assumption that the total density is
constant. Accounting for the inhomogeneity of the total density
should probably lead to an expansion of the range of concentrations
at which the transition to the inhomogeneous phase is possible.

Using the introduced distribution function of inhomogeneous states,
the contribution of inhomogeneity to the entropy and heat capacity
is calculated. This calculation refines and corrects the
corresponding calculation of the previous work [1]. It is shown that
during the transition from the homogeneous state to the
inhomogeneous state and the transition between two inhomogeneous
phases the heat capacity undergoes a jump, so that these transitions
are second-order phase transitions. When approaching zero
temperature, the contribution to the heat capacity from
inhomogeneity decreases linearly with decreasing temperature. An
estimate is given of the temperature range in which the classical
description is acceptable.



\end{document}